\RequirePackage{ifpdf}
\ifpdf 
\documentclass[pdftex]{sigma}
\else
\documentclass{sigma}
\fi

\newcommand{\pa}{\partial}
\newcommand{\la}{\lambda}
\newcommand{\tr}{\mathrm{tr}}
\newcommand{\A}{\mathbb{A}}
\newcommand{\cA}{\mathcal{A}}
\newcommand{\cB}{\mathcal{B}}
\newcommand{\bt}{\mathbf{t}}

\newcommand{\imag}{\mathrm{i}\,}

\numberwithin{equation}{section}

\begin{document}

\allowdisplaybreaks

\renewcommand{\thefootnote}{$\star$}

\renewcommand{\PaperNumber}{002}

\FirstPageHeading

\ShortArticleName{Multicomponent Burgers and KP Hierarchies}

\ArticleName{Multicomponent Burgers and KP Hierarchies,\\ and Solutions from a Matrix Linear System\footnote{This paper is a contribution to the Proceedings of the XVIIth International Colloquium on Integrable Systems and Quantum Symmetries (June 19--22, 2008, Prague, Czech Republic). The full collection
is available at
\href{http://www.emis.de/journals/SIGMA/ISQS2008.html}{http://www.emis.de/journals/SIGMA/ISQS2008.html}}}

\Author{Aristophanes DIMAKIS~$^\dag$ and Folkert M\"ULLER-HOISSEN~$^\ddag$}

\AuthorNameForHeading{A. Dimakis and F. M\"uller-Hoissen}

\Address{$^\dag$~Department of Financial and Management Engineering, \\
 \hphantom{$^\dag$}~University of the Aegean, 31 Fostini Str., GR-82100 Chios, Greece}
\EmailD{\href{mailto:dimakis@aegean.gr}{dimakis@aegean.gr}}

\Address{$^\ddag$~Max-Planck-Institute for Dynamics and Self-Organization, \\
\hphantom{$^\ddag$}~Bunsenstrasse 10, D-37073 G\"ottingen, Germany}
\EmailD{\href{folkert.mueller-hoissen@ds.mpg.de}{folkert.mueller-hoissen@ds.mpg.de}}

\ArticleDates{Received November 01, 2008, in f\/inal form January 04,
2009; Published online January 08, 2009}

\Abstract{Via a Cole--Hopf transformation, the multicomponent linear heat hierarchy
leads to a multicomponent Burgers hierarchy.
We show in particular that any solution of the latter also solves
a corresponding multicomponent (potential) KP hierarchy. A generalization
of the Cole--Hopf transformation leads to a more general relation between
the multicomponent linear heat hierarchy and the multicomponent
KP hierarchy. From this results a construction of exact solutions of the
latter via a matrix linear system.}


\Classification{37K10; 35Q53}

\renewcommand{\thefootnote}{\arabic{footnote}}
\setcounter{footnote}{0}

\section{Introduction}

The well-known Cole--Hopf transformation $\phi = \psi_x \psi^{-1}$
translates the nonlinear Burgers equation into the linear heat equation
\cite{Hopf50,Cole51,Sawa+Kote74}.
This extends to a corresponding relation between the Burgers hierarchy
and the linear heat hierarchy, and moreover to their matrix generalizations.
The Cole--Hopf transformation also generates solutions of the KP hierarchy from
(invertible) solutions of the linear heat hierarchy \cite{MSS91,GMA94,Guil+Mana96},
and this extends to the corresponding matrix hierarchies \cite{DMH07Burgers,DMH07Wronski,Zenc+Sant08}.
There is also a generalization of the Cole--Hopf transformation that produces
a solution of the (scalar or matrix) KP hierarchy from \emph{two} solutions of the matrix linear
heat hierarchy, connected by an additional relation \cite{DMH07Burgers,DMH07Wronski}.
Furthermore, this includes  a construction of (matrix) KP solutions from a matrix linear
system (see also \cite{DMH06nahier,DMH07Ricc}),
which may be regarded as a f\/inite-dimensional version of the Sato theory \cite{Sato+Sato82,Taka89}.
In this work we extend these results to the multicomponent case. Admittedly, this is not
a very dif\/f\/icult task, on the basis of our previous results \cite{DMH07Burgers,DMH07Ricc}.
We also take this opportunity, however, to present these results in a~concise form and
to add some relevant remarks and examples.

The usual multicomponent KP (mcKP) hierarchy contains subhierarchies that generalize the~matrix
KP hierarchy by modifying it with some constant matrix $B$ dif\/ferent from the identity matrix.
We are particularly interested in such matrix hierarchies dif\/ferent from the ordinary
matrix KP hierarchy. The Davey--Stewartson equation \cite{Dave+Stew74,Anke+Free78}, a two-dimensional
nonlinear Schr\"odinger equation that appeared as a shallow-water limit of the Benney--Roskes
equa\-tion~\mbox{\cite{Benn+Rosk69,NakaA82JMP}}, emerges from such a modif\/ied matrix hierarchy
(see Section~\ref{section:mcKP}).
It is of course well-known to arise from the two-component KP hierarchy.

Section~\ref{section:CH} recalls the Cole--Hopf transformation for a multicomponent
Burgers (mcBurgers) hierarchy and Section~\ref{section:mcKP} reveals relations between
the mcBurgers and the mcKP hierarchy. Section~\ref{section:gCH} presents the
abovementioned kind of generalization of the Cole--Hopf transformation which determines
solutions of the mcKP hierarchy. From this derives a fairly simple method to
construct mcKP solutions. This is the subject of Section~\ref{section:mcKPsol}, which
also presents some examples. All this is closely related to a multicomponent version
of a matrix Riccati hierarchy, as brief\/ly explained in Section~\ref{section:mcRiccati}.
Section~\ref{section:concl} contains some conclusions.

In the following, let $\cA$ be an associative (and typically noncommutative) algebra over
the complex (or real) numbers, with identity element $I$ and supplied with a structure that
allows to def\/ine derivatives with respect to real variables.
Let $\cB$ be a f\/inite set of mutually \emph{commuting} elements of $\cA$.
Although this assumption is suf\/f\/icient to establish the results in this work,
further assumptions should be placed on $\cB$ in particular in order
to diminish redundancy, see the remark at the end of Section~\ref{section:CH}.
With each $B \in \cB$ we associate a sequence of (real or complex) variables
$\bt_B = (t_{B,1}, \ldots, t_{B,n}, \ldots)$. Furthermore, for a
function $f$ of $\{ \bt_B \}_{B \in \cB}$ we def\/ine a \emph{Miwa shift}
with respect to $B \in \cB$ by
$ f_{[\la]_B}(\ldots,\bt_B, \ldots) = f(\ldots,\bt_B + [\la], \ldots)$,
where $[\la] = (\la, \la/2, \la/3, \ldots)$ and $\la$ is an indeterminate.
$f_{t_n}$ denotes the partial derivative of $f$ with
respect to the variable $t_n$.

\section[Cole-Hopf transformation for a multicomponent Burgers hierarchy]{Cole--Hopf transformation\\ for a multicomponent Burgers hierarchy}
\label{section:CH}

Let us consider the multicomponent linear heat hierarchy
\begin{gather}
    \psi_{t_{B,n}} = B^n   \pa^n(\psi)  \qquad
    \forall \, B \in \cB   , \qquad n=1,2,\ldots   ,
          \label{mcheathier}
\end{gather}
where $\pa = \pa_x$ is the operator of partial dif\/ferentiation with
respect to a variable $x$, and $\psi$ has values in $\cA$.
Since $\psi_{t_{B,1}} = B   \psi_x$, this implies the ordinary linear heat hierarchy
$\psi_{t_{B,n}} = \pa_{t_{B,1}}^n(\psi)$.
Any two f\/lows (\ref{mcheathier}) commute as a consequence of our assumptions
for $\cB$ (the elements do not depend on $\{ \bt_B \}_{B \in \cB}$ and commute with each other).
A functional representation\footnote{We use this (not quite satisfactory) term for an equation
that generates a sequence of equations by expansion in powers of an indeterminate.}
of (\ref{mcheathier}) is given by
\begin{gather}
    \la^{-1}(\psi - \psi_{-[\la]_B}) = B   \psi_x
    \qquad  \forall \, B \in \cB   .  \label{func_mcheathier}
\end{gather}

\begin{proposition}
If $\psi$ is an invertible solution of the above multicomponent linear heat
hierarchy, then
\begin{gather}
         \phi = \psi_x   \psi^{-1}  \label{Cole--Hopf}
\end{gather}
solves the multicomponent Burgers (mcBurgers) hierarchy associated with $\cB$, given
by the functional representation
\begin{gather}
     \Omega_B(\phi,\la) = 0  \qquad  \forall \, B \in \cB    ,
                  \label{mcBurgers-hier}
\end{gather}
where
\begin{gather}
  \Omega_B(\phi,\la) := \la^{-1} (\phi - \phi_{-[\la]_B})
     - ( B \phi - \phi_{-[\la]_B} B )   \phi - B   \phi_x   .
          \label{mcBurgers-Omega_B}
\end{gather}
\end{proposition}
\begin{proof}
We have to consider the following system,
\[
    \psi_x = \phi   \psi    , \qquad
    \psi_{-[\la]_B} = (I - \la   B   \phi)   \psi
    \qquad  \forall \, B \in \cB   .
\]
The integrability condition $(\psi_x)_{-[\la]_B} = (\psi_{-[\la]_B})_x$
yields (\ref{mcBurgers-hier}). The further integrability condition
$(\psi_{-[\la]_{B_1}})_{-[\mu]_{B_2}} = (\psi_{-[\mu]_{B_2}})_{-[\la]_{B_1}}$
is
\[
   B_2   \big( \la^{-1} (\phi - \phi_{-[\la]_{B_1}})
       + \phi_{-[\la]_{B_1}} B_1   \phi \big)
 = B_1   \big( \mu^{-1} (\phi - \phi_{-[\mu]_{B_2}})
       + \phi_{-[\mu]_{B_2}}   B_2 \phi \big)   ,
\]
which is satisf\/ied as a consequence of (\ref{mcBurgers-hier}) and
$[B_1,B_2] =0$.
\end{proof}

(\ref{Cole--Hopf}) is a Cole--Hopf transformation.
The f\/irst equation that results from (\ref{mcBurgers-hier}) is
\[
    \phi_{t_{B,1}} = B   \phi_x + [B ,\phi]   \phi   ,
\]
which has been called \emph{$C$-integrable $N$-wave equation} in \cite{Zenc+Sant08}.
If $B=I$, this reduces to $t_{I,1} = x$. But it is a nontrivial
nonlinear equation if $\phi$ does not commute with $B$.
The next equation that results from (\ref{mcBurgers-hier}) is the
(noncommutative) Burgers equation
\[
    \phi_{t_{B,2}} = \phi_{t_{B,1}t_{B,1}}
           + 2  \phi_{t_{B,1}}   B   \phi   .
\]

\begin{remark}
In order to avoid redundancy and to maximally extend
the hierarchies, further conditions have to be imposed on~$\cB$.
In particular, the elements of $\cB$ should be linearly independent, since linear
combinations correspond to linear combinations of hierarchy equations.
But since products of elements of $\cB$ also generate (independent or
redundant) commuting f\/lows, the problem is more subtle.
If the algebra is semisimple, it admits a maximal set of commuting mutually
annihilating idempotents~$E_a$, $a =1, \ldots, N$, hence
$E_a   E_b = \delta_{a,b}   E_a$ for all $a,b =1, \ldots, N$.
Then $\cB = \{ E_a \}_{k=1}^N$ is an optimal choice.
In fact, in the following we do not really address those cases of (non-semisimple) algebras
where such a choice does not exist. Rather it turns out that some more f\/lexibility in the
choice of $\cB$ can be used to obtain certain integrable systems
within this framework in a more direct way, see Section~\ref{section:mcKP}.
\end{remark}

\section{Multicomponent KP and relations with the multicomponent\\ Burgers hierarchy}
\label{section:mcKP}

For $B \in \cB$ let
\begin{gather*}
    \mathcal{E}_B(\la) := I - \la   \left( \omega_B(\la) + B \pa \right)    .
\end{gather*}
The ``discrete'' zero curvature condition
\begin{gather}
    \mathcal{E}_{B_1}(\la)_{-[\mu]_{B_2}}   \mathcal{E}_{B_2}(\mu)
  = \mathcal{E}_{B_2}(\mu)_{-[\la]_{B_1}}   \mathcal{E}_{B_1}(\la)
    \label{EE}
\end{gather}
then leads to the two equations
\begin{gather*}
    \la^{-1} (\omega_{B_2}(\mu) - \omega_{B_2}(\mu)_{-[\la]_{B_1}})
     + \omega_{B_2}(\mu)_{-[\la]_{B_1}}   \omega_{B_1}(\la)
     + B_2   \omega_{B_1}(\la)_x \nonumber  \\
   \qquad{} =  \mu^{-1} (\omega_{B_1}(\la) - \omega_{B_1}(\la)_{-[\mu]_{B_2}})
     + \omega_{B_1}(\la)_{-[\mu]_{B_2}}   \omega_{B_2}(\mu)
     + B_1   \omega_{B_2}(\mu)_x
\end{gather*}
and
\[
   B_2   \omega_{B_1}(\la) - \omega_{B_1}(\la)_{-[\mu]_{B_2}}   B_2
 = B_1   \omega_{B_2}(\mu) - \omega_{B_2}(\mu)_{-[\la]_{B_1}}  B_1    .
\]
The last equation is solved by
\[
     \omega_B(\la) = B   \phi - \phi_{-[\la]_B}   B   ,
\]
and the f\/irst equation can then be written in terms of (\ref{mcBurgers-Omega_B}) as
\begin{gather}
   B_2   \Omega_{B_1}(\phi,\la) - \Omega_{B_1}(\phi,\la)_{-[\mu]_{B_2}} \!  B_2
 = B_1   \Omega_{B_2}(\phi,\mu) - \Omega_{B_2}(\phi,\mu)_{-[\la]_{B_1}} \!  B_1
   \!\!\qquad \forall \, B_1,B_2 \!\in\! \cB   . \!\!\!  \label{funct_mcKP}
\end{gather}
We take this as our def\/ining equations of the (more precisely, \emph{potential})
\emph{multicomponent KP (mcKP) hierarchy} associated with $\cB$.\footnote{See also e.g.
\cite{Sato81,DJKM81TIII,KMS90,Oeve93,Berg+tenK95,Guil+Mana96,DMMMS99,Dick03,Kac+vanderLeur03,Kund+Stra95,Bogd+Kono98}
for dif\/ferent formulations of such a multicomponent KP hierarchy.
We should also mention that the conditions imposed on the
set $\cB$ can be relaxed while keeping the hierarchy property,
see \cite{Kund+Stra95} for example.}

\begin{remark}
Choosing $B_1=B_2=B$ in (\ref{funct_mcKP}) and summing the resulting equation
three times with cyclically permuted indeterminates, leads to
\[
  \sum_{i,j,k=1}^3 \epsilon_{ijk}  \left(
     \la_i^{-1} (\phi - \phi_{-[\la_i]_B}) + \phi_{-[\la_i]_B}   B   \phi
           \right)_{-[\la_j]_B}   B = 0    ,
\]
which is a special case of the functional form of the mcKP hierarchy in \cite{Bogd+Kono98}.
\end{remark}

Let us take a closer look at (\ref{funct_mcKP}) with $B_1=B_2=B$.
Its $\la$-independent part is
\begin{gather*}
   B   \Omega_B(\phi,0) - \Omega_B(\phi,0)_{-[\mu]_B}   B
 = [ B , \Omega_B(\phi,\mu) ]   ,         
\end{gather*}
where $\Omega_B(\phi,0) = \phi_{t_{B,1}} - [B,\phi]   \phi - B   \phi_x$.
To f\/irst order in $\mu$ this gives
\begin{gather}
     B   \phi_{t_{B,1} x}   B
   - \tfrac{1}{2} \{ B , \phi_{t_{B,1} t_{B,1}} \} + \tfrac{1}{2} [B , \phi_{t_{B,2}}]
 = B   \phi_{t_{B,1}}   [B,\phi] - [B,\phi]   \phi_{t_{B,1}}   B   .
     \label{Beq-la0mu1}
\end{gather}

(\ref{funct_mcKP}) is the integrability condition of
\begin{gather}
     \Omega_B(\phi,\la) = B   \theta - \theta_{-[\la]_B}   B
            \qquad    \forall \,  B \in \cB      ,
            \label{mcKP-inhomBurg}
\end{gather}
with a new dependent variable $\theta$. (\ref{mcKP-inhomBurg}) represents the
mcKP hierarchy in terms of \emph{two} dependent variables. By comparison with (\ref{mcBurgers-hier}),
this has the form of an \emph{inhomogeneous} mcBurgers hierarchy.
The following is an immediate consequence\footnote{It was f\/irst noted in \cite{MSS91} that any solution
of the f\/irst two equations of the (scalar) Burgers hierarchy also solves the (scalar potential) KP equation.
In \cite{Chen+Li92,Kono+Stra92} the (f\/irst two) Burgers hierarchy equations have been recovered
via a symmetry constraint from the KP hierarchy and its linear system.}.

\begin{proposition}
Any solution of the mcBurgers hierarchy also solves the mcKP hierarchy.
\end{proposition}
\begin{proof}
(\ref{mcKP-inhomBurg}) becomes (\ref{mcBurgers-hier}) if $\theta=0$.
\end{proof}

The representation (\ref{mcKP-inhomBurg}) of the mcKP hierarchy has the advantage that each equation
only involves a single element from the set $\cB$.
To order $\la^0$, (\ref{mcKP-inhomBurg}) yields
\begin{gather}
    \phi_{t_{B,1}} - B   \phi_x - [B,\phi]   \phi = [B,\theta]   .
         \label{mcKP-inhomBurg1}
\end{gather}
If $B=I$, then (\ref{mcKP-inhomBurg1}) reduces to
$\phi_{t_{B,1}} = \phi_x$. Otherwise this is a non-trivial nonlinear equation.
Subtracting (\ref{mcKP-inhomBurg1}) from (\ref{mcKP-inhomBurg}), leads to
\begin{gather}
   \la^{-1} (\phi - \phi_{-[\la]_B}) - \phi_{t_{B,1}}
     - (\phi - \phi_{-[\la]_B})   B   \phi = (\theta - \theta_{-[\la]_B})   B   .
         \label{mcKP-inhomBurg2}
\end{gather}
The two equations (\ref{mcKP-inhomBurg1}) and (\ref{mcKP-inhomBurg2})
are equivalent to (\ref{mcKP-inhomBurg}). (\ref{mcKP-inhomBurg2}) does not involve
derivatives with respect to $x$. For f\/ixed $B \in \cB$, it represents the
KP hierarchy in $\cA$ \cite{DMH07Burgers}, with product modif\/ied by $B$.

\begin{proposition}
For any $B \in \cB$, as a consequence of the mcKP hierarchy \eqref{mcKP-inhomBurg},
$B \phi$ and also $\phi B$ solves the ordinary (noncommutative) KP hierarchy.
\end{proposition}
\begin{proof}
This is an immediate consequence of (\ref{mcKP-inhomBurg2}).
\end{proof}

To f\/irst order in $\la^{-1}$, (\ref{mcKP-inhomBurg2}) yields
\[
     \tfrac{1}{2} (\phi_{t_{B,2}} - \phi_{t_{B,1}t_{B,1}} )
   - \phi_{t_{B,1}}   B   \phi = \theta_{t_{B,1}}   B   .
\]
Dif\/ferentiating (\ref{mcKP-inhomBurg1}) with respect to $t_{B,1}$,
and multiplying it by $B$ from the right, we can use the last equation
to eliminate $\theta$. In this way we recover (\ref{Beq-la0mu1}).
As shown in Example~\ref{ex:ds} below, (\ref{Beq-la0mu1})~generalizes
the \emph{Davey--Stewartson} (DS) equation \cite{Benn+Rosk69,Dave+Stew74,Anke+Free78,NakaA82JMP}.
Eliminating $\theta$ from (\ref{mcKP-inhomBurg2}) by use of
(\ref{mcKP-inhomBurg1}) thus leads to a (generalized) DS hierarchy.

\begin{example}
\label{ex:2cKP}
Let $B = \sigma$, where $\sigma^2 = I$. Decomposing $\phi$ as
\[
    \phi = D + U, \qquad \mbox{where} \qquad
       D := \tfrac{1}{2}(\phi + \sigma   \phi   \sigma)   , \qquad
       U := \tfrac{1}{2}(\phi - \sigma   \phi   \sigma)   ,
\]
(\ref{Beq-la0mu1}) splits into the two equations\footnote{A constant (with respect to $t_1$)
of integration has been set to zero in order to obtain (\ref{DS_diag}). The latter equation
can be obtained more directly as the diagonal part of (\ref{mcKP-inhomBurg1}).}
\begin{gather}
    D_{t_1}   \sigma - D_x - 2   U^2  =  0,     \label{DS_diag}
  \\
    U_{t_2}  \sigma + U_{x t_1} + 2  \{ U , D_{t_1} \} = 0  ,
                                      \label{DS_offd}
\end{gather}
where we write $t_1$, $t_2$ instead of $t_{B,1}$ and $t_{B,2}$.
Let now $\cA$ be the algebra of $2 \times 2$ matrices over~$\mathbb{C}$,
and $\sigma = \mathrm{diag}(1,-1)$. Then $D$ and $U$ are diagonal and of\/f-diagonal parts
of $\phi$, respectively. Writing
\begin{gather}
    U = \left(\begin{array}{cc} 0 & u \\ v & 0 \end{array}\right)   ,
        \qquad
   \mbox{hence} \qquad u := \phi_{1,2}   , \qquad v := \phi_{2,1}   ,
    \label{U->uv}
\end{gather}
and introducing
\[
    s := \tr(\phi) = \phi_{1,1} + \phi_{2,2}   , \qquad
    r := \phi_{1,1} - \phi_{2,2}   ,
\]
we obtain the system
\begin{gather}
   u_{t_2} - u_{xt_1} - 2   u   s_{t_1} = 0   , \qquad
   v_{t_2} + v_{xt_1} + 2   v   s_{t_1} = 0   ,   \label{ex1_u,v_eqs}
\end{gather}
and $s_{t_1} = r_x$, $s_x = r_{t_1} - 4   u   v$.
The integrability conditions of the latter two equations are
\begin{gather}
   s_{t_1t_1} - s_{xx} = 4   (u   v)_x   , \qquad
   r_{t_1t_1} - r_{xx} = 4   (u   v)_{t_1}    .   \label{ex1_s,r_eqs}
\end{gather}
The two equations (\ref{ex1_u,v_eqs}) together with the f\/irst of (\ref{ex1_s,r_eqs})
constitute a fairly simple system of three nonlinear coupled equations, where all
variables can be taken to be real.
Allowing complex dependent and independent variables, after a complex transformation
the system for the dependent variables $u$, $v$, $s$ can be further reduced to the DS
equation, see the next example.
Of course, the above system (\ref{ex1_u,v_eqs}) and (\ref{ex1_s,r_eqs}) can also be
derived from the usual two-component KP hierarchy (see e.g.~\cite{Bogd+Kono98}), and the
transformation to DS is well-known.

Setting $v=0$, we obtain from (\ref{ex1_u,v_eqs}) and (\ref{ex1_s,r_eqs}) the following
\emph{linear} equations\footnote{If $u$ does not depend on $t_2$, then the f\/irst equation
is part of a Lax pair for the Nizhnik--Novikov--Veselov equation. },
\begin{gather}
   u_{t_2} - u_{xt_1} - 2   \varphi   u = 0   , \qquad
   \varphi_{t_1t_1} - \varphi_{xx} = 0   ,
   \label{u,s_eqs_v=0}
\end{gather}
where $\varphi := s_{t_1} = \tr(\phi)_{t_1}$.
This is probably the simplest system that possesses \emph{dromion} solutions, as observed
in Example~\ref{ex:c1-ex1} in Section~\ref{section:mcKPsol}.
\end{example}

\begin{example}
\label{ex:ds}
We continue with the previous example and perform a transformation to the
DS equation by f\/irst allowing the dependent variables to live in a
\emph{non}commutative algebra. In this way we obtain a certain noncommutative
generalization of the DS equation (see also \cite{Yuro96,Lezn+Yuzb97} for matrix DS versions).
In terms of
\[
    F = D_{t_1} + \beta   U^2   ,
\]
with $\beta \in \mathbb{C}$, (\ref{DS_diag}) becomes
\[
    F_{t_1}   \sigma - F_x = - \beta   (U^2)_x
      + (U^2)_{t_1}   (2 + \beta   \sigma)  .
\]
Dif\/ferentiating this with respect to $x$ and with respect to $t_1$,
respectively, and eliminating mixed derivatives of $F$ from the resulting
two equations, yields
\begin{gather}
    F_{t_1 t_1} - F_{x x} = - \beta   (U^2)_{x x} + 2   (U^2)_{x t_1}
                        + (U^2)_{t_1 t_1}   (2 \sigma + \beta)   .
       \label{DS_F-eq}
\end{gather}
 Furthermore, (\ref{DS_offd}) takes the form
\begin{gather}
    U_{t_2}   \sigma + U_{x t_1} = - 2   \{ U , F \} + 4   \beta   U^3   .
        \label{DS_U-eq}
\end{gather}
Let now $\cA$ be the algebra of $2 \times 2$ matrices over
some unital associative algebra with unit denoted by~$1$, and
$\sigma = \mathrm{diag}(1,-1)$. Using (\ref{U->uv}) and writing
\[
    F = \mathrm{diag}(f , g)   ,
\]
(\ref{DS_F-eq}) and (\ref{DS_U-eq}) result in the following equations,
\begin{gather*}
    u_{t_2} - u_{x t_1} = 2   (f u + u g) - 4   \beta u v u   , \nonumber \\
    v_{t_2} + v_{x t_1} = -2   (g v + v f) + 4 \, \beta v u v   , \nonumber \\
    f_{t_1 t_1} - f_{x x} = - \beta   (u v)_{x x} + 2   (u v)_{x t_1} + (2 + \beta)   (u v)_{t_1 t_1}
                          , \nonumber \\
    g_{t_1 t_1} - g_{x x} = - \beta   (v u)_{x x} + 2   (v u)_{x t_1} - (2 - \beta)   (v u)_{t_1 t_1} .
\end{gather*}
In terms of the new variables
\begin{gather}
    y = \frac{1+\imag}{\kappa   \sqrt{2}} (x + \imag t_1)   , \qquad
    z = \frac{1-\imag}{\sqrt{2}} (x - \imag t_1)   ,  \label{DS_transf}
\end{gather}
with a constant $\kappa \neq 0$,
and with the choice $\beta = \imag$, this becomes
\begin{gather}
        u_{t_2} + u_{zz} + \frac{1}{\kappa^2} u_{y y}
= 2  (f u +u g) - 4   \imag u v u   , \nonumber \\
      - v_{t_2} + v_{zz} + \frac{1}{\kappa^2} v_{y y}
= 2(g v+v f) + 4   \imag v u v   , \nonumber \\
        f_{zz} - \frac{1}{\kappa^2} f_{y y}
= (1+2 \imag)(u v)_{zz} - \frac{1}{\kappa^2} (u v)_{y y}
        - \frac{2 \imag}{\kappa} (u v)_{y z}   , \nonumber \\
        g_{zz} - \frac{1}{\kappa^2} g_{y y}
= (-1 + 2 \imag) (v u)_{zz} + \frac{1}{\kappa^2} (v u)_{y y}
        + \frac{2 \imag}{\kappa} (v u)_{y z}   .  \label{ncDS}
\end{gather}
If the dependent variables take their values in a \emph{commutative} algebra,
then we obtain
\begin{gather*}
      u_{t_2} + u_{z z} + \frac{1}{\kappa^2} u_{y y}
  = 2   \rho   u - 4   \imag u^2 v   , \nonumber \\
     -v_{t_2} + v_{z z} + \frac{1}{\kappa^2} v_{y y}
  = 2   \rho  v + 4   \imag v^2 u   , \nonumber \\
      \rho_{zz} - \frac{1}{\kappa^2} \rho_{y y}
  = 4  \imag   (u v)_{zz}   ,
\end{gather*}
where
\begin{gather}
   \rho := f+g = \varphi + 2   \imag   u   v
              \label{DS_rho}
\end{gather}
(with $\varphi = \tr(\phi)_{t_1}$). Setting $t_2 = -\imag t$ and
\[
      v = \imag \epsilon   \bar{u} \qquad
       \mbox{with} \qquad  \epsilon = \pm 1
\]
and the complex conjugate $\bar{u}$ of $u$, one recovers the DS system
\begin{gather}
     \imag   u_t + u_{z z} + \frac{1}{\kappa^2} u_{y y}
   = 2   \rho   u + 4   \epsilon  |u|^2 u   ,  \qquad
      \rho_{z z} - \frac{1}{\kappa^2} \rho_{y y}
    = - 4  \epsilon   (|u|^2)_{zz}   ,  \label{DS}
\end{gather}
where now $y$, $z$, $t$, $\kappa^2$, $\rho$ are taken to be real. Clearly, this system
is more quickly obtained from~(\ref{ex1_u,v_eqs}) and (\ref{ex1_s,r_eqs})
by application of the transformation of independent variables given by~(\ref{DS_transf}) and $t_2 = -\imag t$, and the transformation (\ref{DS_rho})
of dependent variables.
But on our way we obtained the system (\ref{ncDS}) which (with $t_2 = -\imag t$)
may be of interest as a noncommutative version of the DS equation (see
\cite{Yuro96,Lezn+Yuzb97} for alternatives).
\end{example}

We conclude that the DS equation (and a corresponding hierarchy) is obtained from a
\emph{matrix} KP hierarchy, but the latter has to be generalized by introduction of
a matrix $B$ dif\/ferent from the unit matrix.
Of course, this hierarchy is embedded in the usual two-component KP hierarchy.

\subsection[The associated Sato-Wilson system and its translation into
a Burgers hierarchy]{The associated Sato--Wilson system and its translation\\ into
a Burgers hierarchy}

(\ref{EE}) is the integrability condition of the linear system
\[
    \tilde{\psi}_{-[\la]_B} = \mathcal{E}_B(\la)   \tilde{\psi}
    \qquad \forall \, B \in \cB   .
\]
If $W$ is an invertible solution of
\begin{gather}
    \mathcal{E}_B(\la)   W = W_{-[\la]_B}   \mathcal{E}_{0,B}(\la)   ,
    \qquad \mathcal{E}_{0,B}(\la) := I - \la   B   \pa   ,
             \label{Sato-Wilson}
\end{gather}
then the linear system is mapped to
$\psi_{-[\la]_B} = \mathcal{E}_{0,B}(\la)   \psi$ where $\psi := W^{-1}   \tilde{\psi}$.
The latter is the linear heat hierarchy (\ref{func_mcheathier}). The ansatz
\[
   W = I + \sum_{n \geq 1} w_n  \pa^{-n}   ,
\]
inserted into the functional form (\ref{Sato-Wilson}) of the Sato--Wilson equations,
leads to
\begin{gather}
 \la^{-1}(w_n - w_{n,-[\la]_B}) - B w_{n,x} -(B \phi-\phi_{-[\la]_B} B)   w_n
 = B w_{n+1} - w_{n+1,-[\la]_B} B   ,   \label{Sato-Wilson2}
\end{gather}
where $n=0,1,\ldots$ and $w_0=I$. From the $n=0$ equation we get $w_1=-\phi$.
In terms of $\theta := -w_2$, the $n=1$ equation turns out to be
the functional form (\ref{mcKP-inhomBurg}) of the mcKP hierarchy. Let us introduce
\begin{gather}
    \Phi := \Lambda^\intercal + e_1   (\phi,-w_2,-w_3,\ldots)   ,
                         \label{Phi=LaT+e_1(phi...)}
\end{gather}
where $e_1^\intercal = (I,0,\ldots)$, and
\begin{gather}
     \Lambda := \left( \begin{array}{ccccc}
               0 &   I       &          0 & 0      & \cdots \\
               0 &   0         &  I       & 0      & \cdots \\
          \vdots &  \vdots     &  \ddots    & \ddots & \ddots
                 \end{array} \right)   \label{Lambda}
\end{gather}
is the shift operator matrix with transpose $\Lambda^\intercal$. Then
(\ref{Sato-Wilson2}) can be expressed as the following mcBurgers hierarchy,
\begin{gather}
  \Omega_B(\Phi,\la) \equiv \la^{-1} (\Phi - \Phi_{-[\la]_B})
     - ( B \Phi - \Phi_{-[\la]_B} B )   \Phi - B   \Phi_x
   = 0  \qquad \forall \, B \in \cB    .   \label{inf_mcBurgers}
\end{gather}
The linear system of the mcKP hierarchy has thus been reformulated as an mcBurgers
hierarchy with an inf\/inite matrix variable of a special form.

We note that (\ref{Phi=LaT+e_1(phi...)}) has the form of a
\emph{companion matrix}, a block of a \emph{Frobenius normal form} matrix.
This makes contact with recent work in \cite{Zenc+Sant08}, where several
integrable systems were recovered from equations of a Burgers hierarchy
with dependent variable of Frobenius normal form. All these systems are
known to arise as reductions of the mcKP hierarchy.

As a consequence of (\ref{inf_mcBurgers}) with $\Phi$ of the form
(\ref{Phi=LaT+e_1(phi...)}), it follows that $\phi = e_1^\intercal \Phi e_1$
solves the mcKP hierarchy in $\cA$.

In the following section, we shall see that $\Phi$ of the form (\ref{Phi=LaT+e_1(phi...)})
results, as a particular case, from a quite general result about solutions of
a somewhat generalized mcKP hierarchy.

\section[Generalization of the Cole-Hopf transformation]{Generalization of the Cole--Hopf transformation}
\label{section:gCH}

We generalize (\ref{mcKP-inhomBurg}) to
\begin{gather}
   \Omega_{B,Q}(\Phi,\la) = B   \Theta - \Theta_{-[\la]_B} B
       \qquad \forall \, B \in \cB   ,
          \label{funct_mcKP_Q}
\end{gather}
where
\begin{gather}
  \Omega_{B,Q}(\Phi,\la)
 := \la^{-1}(\Phi - \Phi_{-[\la]_B})
   - ( B \Phi - \Phi_{-[\la]_B} B )   Q   \Phi - B   \Phi_x   ,
     \label{Omega_BQ}
\end{gather}
with a constant object $Q$.
We assume that the constituents are elements of linear spaces and that
the products are def\/ined. (\ref{funct_mcKP_Q}) determines a generalization
of the mcKP hierarchy, since nonlinear terms now involve $Q$ (which modif\/ies
the product).
The following generalizes a theorem in~\cite{DMH07Burgers} (see also~\cite{DMH07Wronski}).

\begin{theorem}
\label{theorem:gCH}
Let $X$, $Y$ be solutions of the multicomponent linear heat hierarchy, i.e.
\begin{gather}
   \la^{-1}(X - X_{-[\la]_B}) = B   X_x    ,  \qquad
   \la^{-1}(Y - Y_{-[\la]_B}) = B   Y_x   ,  \label{X,Y-mcheathier}
\end{gather}
for all $B \in \cB$, and furthermore
\begin{gather}
    X_x = R   X + Q   Y   ,  \label{Xx=RX+QY}
\end{gather}
with constant objects $R$, $Q$. If $X$ is invertible and if all $B \in \cB$
commute with $R$, then
\begin{gather}
    \Phi = Y X^{-1}    \label{Phi=YX^-1}
\end{gather}
solves the mcKP$_Q$ hierarchy \eqref{funct_mcKP_Q} with $\Theta = \Phi R$.
\end{theorem}

\begin{proof}
Using (\ref{Phi=YX^-1}) in the def\/inition
(\ref{Omega_BQ}), we have
\begin{gather*}
   \Omega_{B,Q}(\Phi,\la)
 = (B \Phi - \Phi_{-[\la]_B} B)(X_x - Q Y) X^{-1}
     +( \la^{-1} (Y - Y_{-[\la]_B}) - B Y_x) X^{-1} \\
\hphantom{\Omega_{B,Q}(\Phi,\la)=}{}  - \Phi_{-[\la]_B} ( \la^{-1} (X - X_{-[\la]_B}) - B X_x) X^{-1}   ,
\end{gather*}
which reduces to
\[
   \Omega_{B,Q}(\Phi,\la) = (B \Phi - \Phi_{-[\la]_B} B)   R
\]
as a consequence of (\ref{X,Y-mcheathier}) and (\ref{Xx=RX+QY}).
Since $[B , R] =0$, this takes the form (\ref{funct_mcKP_Q})
with $\Theta = \Phi R$.
\end{proof}

Now we set up the stage for applications of the theorem.
Let $\A(M,N) := \mathrm{Mat}(M \times N, \mathbb{C}) \otimes \cA$, where
$\mathrm{Mat}(M \times N, \mathbb{C})$ is the space of complex $M \times N$ matrices.
Let $\Phi$, $\Theta$, $Y$ take values in $\A(M,N)$, and $X$ in $\A(N,N)$. Furthermore,
let $Q \in \A(N,M)$ and $R \in \A(N,N)$ commute with all $B \in \cB$.

If $Q$ has \emph{rank one over} $\cA$, in the sense that $Q = V U^\intercal$ with constant
vectors $U$, $V$, with entries in $\cA$, and if $U$ and $V$ commute with all $B \in \cB$,
then $\phi = U^\intercal \Phi V$ solves the mcKP hierarchy
in $\cA$, provided that $\Phi$ solves (\ref{funct_mcKP_Q}).
In this way, any solution $X$, $Y$ of the linear equations formulated in the
above theorem generates an $\cA$-valued solution $\phi$ of the mcKP hierarchy
(\ref{funct_mcKP}).

Choosing $M=N$, (\ref{Phi=YX^-1}) is a \emph{Cole--Hopf} transformation if
$Y = X_x$. Then (\ref{Xx=RX+QY}) becomes $(I-Q) \, X_x = R \, X$.
Let $\mathbf{e}_k$ denote the $N$-component vector with all entries zero
except for the identity element in the $k$th row.

\textbf{a)} Setting $Q = \mathbf{e}_N \mathbf{e}_N^\intercal$ and
$R = \sum\limits_{k=1}^{N-1} \mathbf{e}_k \mathbf{e}_{k+1}^\intercal$ (which is the
left shift operator: $R \mathbf{e}_k = \mathbf{e}_{k-1}$, $k=2,\ldots,N$,
and $R \mathbf{e}_1 =0$), one f\/inds that (\ref{Xx=RX+QY}) restricts $X$ to
the form of a \emph{Wronski matrix} (see also~\cite{DMH07Wronski}),
\[
 X = \left(\begin{array}{cccc} X^{(1)} & X^{(2)} & \cdots & X^{(N)} \\
    \pa(X^{(1)}) &\pa(X^{(2)}) & \cdots & \pa(X^{(N)}) \\
    \vdots & \vdots & \ddots& \vdots \\
    \pa^{N-1}(X^{(1)}) & \pa^{N-1}(X^{(2)}) & \cdots & \pa^{N-1}(X^{(N)})
          \end{array}\right)  .
\]
$X^{(1)}, X^{(2)}, \ldots, X^{(N)}$ are independent functions with values in $\cA$.
The remaining assumption~(\ref{X,Y-mcheathier}) in the theorem requires them to be
solutions of the multicomponent heat hierarchy.

\textbf{b)} Let $R=\Lambda^\intercal$ with the inf\/inite shift operator matrix
(\ref{Lambda}), and $Q=\mathbf{e}_1 \mathbf{e}_1^\intercal$. Then
(\ref{Xx=RX+QY}) says that $X$ has to be a \emph{pseudo-Wronski matrix}
\begin{gather}
    X = \left( \begin{array}{cccc}
                 X^{(1)} & X^{(2)} & X^{(3)} & \cdots \\
               \pa^{-1}X^{(1)} & \pa^{-1}X^{(2)} & \pa^{-1}X^{(3)} & \cdots \\
               \pa^{-2}X^{(1)} & \pa^{-2}X^{(2)} & \pa^{-2}X^{(3)} & \cdots \\
                  \vdots    &  \vdots  & \vdots     & \ddots
                 \end{array} \right)   ,  \label{X-pseudoWronski}
\end{gather}
where $\pa^{-1}$ is the formal inverse of $\pa$. This structure appeared
in \cite{Zenc+Sant08} (equations (14) and (50) therein). (\ref{X,Y-mcheathier})
demands that $X^{(1)}, X^{(2)}, \ldots$ solve the multicomponent heat hierarchy.
With~(\ref{X-pseudoWronski}), $\Phi = X_x X^{-1}$ has the form
(\ref{Phi=LaT+e_1(phi...)}) and hence determines a solution of (\ref{inf_mcBurgers}).

\section{Solutions of the multicomponent KP hierarchy\\ from a matrix linear system}
\label{section:mcKPsol}

In order to derive some classes of mcKP solutions via theorem~\ref{theorem:gCH} more explicitly,
in the framework specif\/ied in Section~\ref{section:gCH} (after the theorem) we extend (\ref{Xx=RX+QY}) to
\begin{gather}
    Z_x = H   Z   ,  \label{Z_x=HZ}
\end{gather}
where
\[
    Z = \left(\begin{array}{c} X \\ Y \end{array}\right)   , \qquad
    H = \left(\begin{array}{cc} R & Q \\ S & L \end{array}\right)   ,
\]
with new constant objects $L \in \A(M,M)$ and $S \in \A(M,N)$
that commute with all $B \in \cB$.
The two equations (\ref{X,Y-mcheathier}) then combine to
\[
    \la^{-1}(Z - Z_{-[\la]_B}) = B   Z_x  \qquad   \forall \, B \in \cB    .
\]
Taking (\ref{Z_x=HZ}) into account, this is equivalent to
\begin{gather}
     Z_{t_{B,n}} = (B H)^n Z   \qquad \quad \forall \, B \in \cB   ,
       \qquad n=1,2,\ldots  .  \label{Z_tBn=(BH)^nZ}
\end{gather}
Note that $B$ and $H$ commute as a consequence of our assumptions.
With a suitable choice of the algebra $\cA$, the solution of the matrix linear system
(\ref{Z_x=HZ}), (\ref{Z_tBn=(BH)^nZ}) is given by
\begin{gather}
    Z(x,\mathbf{t}) = e^{\xi(x,\mathbf{t};H,B)}   Z_0   ,  \label{Zsol}
\end{gather}
where
\[
    \xi(x,\mathbf{t};H,B) := x   H
     + \sum_{B \in \cB} \sum_{n=1}^\infty t_{B,n}   (H B)^n   ,
\]
and $\mathbf{t}$ stands for $\{ \mathbf{t}_B \}_{B \in \cB}$.
Decomposing $Z$ into $X$ and $Y$, theorem~\ref{theorem:gCH} implies that $\Phi = Y X^{-1}$
solves the mcKP$_Q$ hierarchy (\ref{funct_mcKP_Q}). Furthermore,
if $\mathrm{rank}(Q)=1$ over $\cA$, hence $Q = V U^\intercal$ with constant
vectors $U$ and $V$, then the $\cA$-valued variable
\begin{gather*}
       \phi = U^\intercal \Phi V  
\end{gather*}
solves the corresponding mcKP hierarchy.

The exponential in (\ref{Zsol}) can be computed explicitly if additional assumptions are
made concerning the form of $H$ (see~\cite{DMH07Ricc}, in particular). Then $\Phi$ is
obtained via~(\ref{Phi=YX^-1}).

\medskip

\noindent
\textbf{Case 1.} Let $S=0$ and
\begin{gather}
     Q = R K - K L    \label{Q=RK-KL}
\end{gather}
with a constant $N \times M$ matrix $K$ (over $\cA$) that commutes with all $B \in \cB$. Then we obtain
\begin{gather}
   \Phi = e^{\xi(x,\mathbf{t};L,B)}   C   e^{-\xi(x,\mathbf{t};R,B)}
        \big(I_N - K   e^{\xi(x,\mathbf{t};L,B)}   C   e^{-\xi(x,\mathbf{t};R,B)} \big)^{-1}   ,
        \label{Phi-type1}
\end{gather}
where $I_N$ is the $N \times N$ unit matrix over $\cA$ (so that the diagonal entries are
the identity $I$ in~$\cA$), and $C$ is an arbitrary constant $M \times N$ matrix (with entries in $\cA$).
$\Phi$ solves the mcKP$_Q$ hierarchy (associated with $\cB$),
with $Q$ given by (\ref{Q=RK-KL}).\footnote{In particular, if $M=N$ and $Q=I_N$, then $\Phi$
(with $K$, $L$, $R$ solving $R K - K L = I_N$) is a solution of the $N \times N$ matrix (over $\cA$)
mcKP hierarchy.}
If moreover $Q = V U^\intercal$ with vectors $U$, $V$ that commute with all $B \in \cB$, then
$\phi = U^\intercal \Phi V$ solves the mcKP hierarchy in $\cA$.
Of course, it remains to solve the rank one condition (over $\cA$)
\begin{gather}
          R K - K L = V U^\intercal   .   \label{RK-KL=VU^T}
\end{gather}
If $M=N$ and if $C$ is invertible, then (\ref{Phi-type1}) simplif\/ies to
\begin{gather*}
   \Phi = \big(e^{\xi(x,\mathbf{t};R,B)}   C^{-1}   e^{-\xi(x,\mathbf{t};L,B)} - K \big)^{-1}   ,
\end{gather*}
which remains a solution if we replace $C^{-1}$ by an arbitrary constant $N \times N$ matrix $\tilde{C}$
(with entries in $\cA$).

\begin{example}
\label{ex:c1-ex1}
Choosing the components of the matrices $L$, $R$ as
\begin{gather}
    L_{ij} = p_i   \delta_{ij}   I   , \qquad
    R_{kl} = q_k   \delta_{kl}   I   ,    \label{case1_L,R}
\end{gather}
with constants $p_i$, $q_k$, (\ref{RK-KL=VU^T}) is solved by
\begin{gather}
    K_{kj} = \frac{1}{q_k-p_j}   u_k   v_j   ,   \label{case1_ex_K}
\end{gather}
where $u_k$ and $v_j$ are the components of $U$ and $V$, respectively. We elaborate one of the
simplest cases in some detail.
Let us choose $\cA$ as the algebra of $2 \times 2$ matrices over $\mathbb{C}$,
$L = p   I_2$, $R = q   I_2$, with constants $p$ and $q$, $Q=I=I_2$, and $\cB = \{ B \}$
with $B = \mathrm{diag}(1,-1)$ (motivated by the examples in Section~\ref{section:mcKP}).
Then we have
$\xi(x,\mathbf{t};L,B) = \mathrm{diag}( \xi_+(x,\mathbf{t};p) , \xi_-(x,\mathbf{t};p) )$ with
\begin{gather}
    \xi_{\pm}(x,\mathbf{t};p) := p   x + \sum_{n=1}^\infty p^{2n}   t_{2n}
       \pm  \sum_{n=0}^\infty p^{2n+1}   t_{2n+1}   \label{xi_pm}
\end{gather}
(writing $t_n$ instead of $t_{B,n}$),
and $\xi(x,\mathbf{t};R,B)$ is obtained by exchanging $p$ by $q$ in these expressions. Writing
\begin{gather}
     C = \left( \begin{array}{cc} c_1 & c_2 \\ c_3 & c_4 \end{array} \right)   \label{Phi0}
\end{gather}
with constants $c_i$, we obtain (with $U=V=I_2$)\footnote{Since
$\xi_\pm(x,\mathbf{t};p)-\xi_\pm(x,\mathbf{t};q) = (p-q)   (x \pm t_1) + (p^2 -q^2)   t_2 + \cdots$,
this solution becomes independent of $t_2$ (i.e.\ ``static'') if $p=-q$.}
\begin{gather}
  \phi(x,\mathbf{t}) = \frac{1}{\mathcal{D}(x,\mathbf{t})} \left( \begin{array}{cc}
   c_1   e^{\xi_+(x,\mathbf{t};p)-\xi_+(x,\mathbf{t};q)} + f(x,\mathbf{t})  &
                 c_2   e^{\xi_+(x,\mathbf{t};p)-\xi_-(x,\mathbf{t};q)}  \\
   c_3   e^{\xi_-(x,\mathbf{t};p)-\xi_+(x,\mathbf{t};q)}  &
   c_4   e^{\xi_-(x,\mathbf{t};p)-\xi_-(x,\mathbf{t};q)} + f(x,\mathbf{t})
           \end{array} \right) ,    \label{c1ex1_phi}
\end{gather}
where
\begin{gather*}
     f(x,\mathbf{t}) := \frac{c_1 c_4 - c_2 c_3}{p-q}\,   e^{\xi_+(x,\mathbf{t};p) + \xi_-(x,\mathbf{t};p)
                               - \xi_+(x,\mathbf{t};q) - \xi_-(x,\mathbf{t};q)}    ,  \\
     \mathcal{D}(x,\mathbf{t}) := 1 + \frac{1}{p-q}   \big( c_1   e^{\xi_+(x,\mathbf{t};p)
      - \xi_+(x,\mathbf{t};q)} + c_4   e^{\xi_-(x,\mathbf{t};p) - \xi_-(x,\mathbf{t};q)}
      + f(x,\mathbf{t}) \big) .
\end{gather*}
This is a solution of (\ref{Beq-la0mu1}), with $B = \mathrm{diag}(1,-1)$, and its hierarchy,
and its components thus provide us with the solution
\begin{gather}
   u =\frac{c_2}{\mathcal{D}} \, e^{\xi_+(x,\mathbf{t};p)-\xi_-(x,\mathbf{t};q)}   , \qquad
   v = \frac{c_3}{\mathcal{D}} \, e^{\xi_-(x,\mathbf{t};p)-\xi_+(x,\mathbf{t};q)}   , \nonumber \\
   s = \frac{1}{\mathcal{D}}   \big( c_1   e^{\xi_+(x,\mathbf{t};p)-\xi_+(x,\mathbf{t};q)}
         + c_4 \, e^{\xi_-(x,\mathbf{t};p)-\xi_-(x,\mathbf{t};q)} + 2   f \big)
              \label{exc1_u,v,s}
\end{gather}
of the system (\ref{ex1_u,v_eqs}), (\ref{ex1_s,r_eqs}).
$\phi$ is \emph{regular} (for all $\mathbf{t}$) in particular if all constants are real, $c_1 c_4 > c_2 c_3$,
and either $p > q$, $c_1 >0$, $c_4 >0$, or $p < q$, $c_1 <0$, $c_4 <0$.
Fig.~\ref{fig:dromion} presents a~dromion\footnote{The characteristic features of a dromion are
its exponential localization and that it is accompanied by a f\/ield structure of intersecting line solitons.
See \cite{BLMP88,Foka+Sant89,Sant90,Hiet+Hiro90,JMM90,Nakao+Wada93,Sall94,Gils+Nimm91,Peli95DS,Guil+Mana95}
for the Davey--Stewartson case, and especially \cite{Hiet90} for an illuminating structural analysis
and the appearance of dromions as solutions of other equations.}
solution within this family.

\begin{figure}[t]
\centerline{\includegraphics[width=13cm]{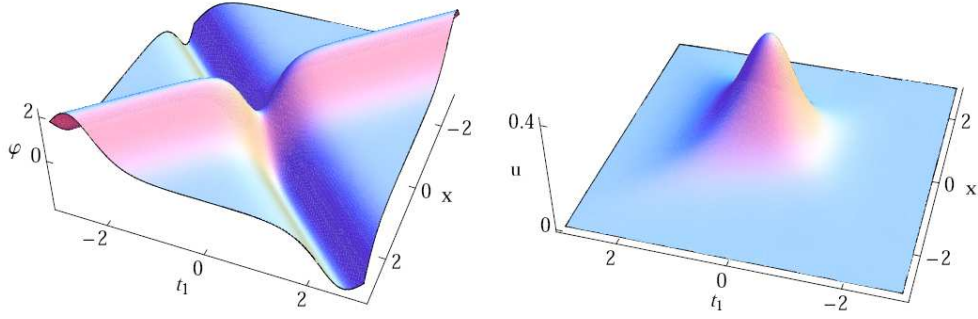}}

\caption{A dromion solution of (\ref{ex1_u,v_eqs}) and (\ref{ex1_s,r_eqs}) at $t_2=0$,
given by (\ref{exc1_u,v,s}) with $p=2$, $q=-1$, $c_1=c_4=1$, $c_2=1/2$ and $c_3=0$.
The left plot shows $\varphi = s_{t_1} = \tr(\phi)_{t_1}$.
As a consequence of $c_3=0$, we have $v=0$. Hence this is actually a solution of the
\emph{linear} equations (\ref{u,s_eqs_v=0}) and $\varphi$ solves the linear wave equation.
Regarding $t_1$ as an evolution parameter, the plot of $\varphi$ shows two colliding humps
(with amplitudes having opposite signs) that annihilate at $t_1=0$.
With $0 \neq c_3 < 2$, the plots remain qualitatively the same as long as $c_3$ is
suf\/f\/iciently far below the upper bound, and $v$ attains a shape similar to that of $u$.
   \label{fig:dromion} }
\end{figure}

If $c_3=0$, we have $v=0$ and (\ref{exc1_u,v,s}) determines a solution of the \emph{linear} equations
(\ref{u,s_eqs_v=0}).\footnote{In this case, $\mathcal{D}$ factorizes,
\[
    \mathcal{D} = \left( 1+\frac{c_1}{p-q} \, e^{\xi_+(x,\mathbf{t};p)-\xi_+(x,\mathbf{t};q)} \right)
        \left( 1+\frac{c_4}{p-q} \, e^{\xi_-(x,\mathbf{t};p)-\xi_-(x,\mathbf{t};q)} \right)   .
\] }
An extremum of $u$ for a regular solution then moves (in ``time'' $t_2$) with constant amplitude
along the curve given by{\samepage
\[
   x = -(p+q)   t_2 - \frac{1}{2 (p-q)}   \log\left( \frac{c_1   c_4}{(p-q)^2} \right)   , \qquad
   t_1 = \frac{1}{p-q}   \log\left( -\frac{p}{q}   \sqrt{\frac{c_4}{c_1}} \right)   .
\]
The last expression shows that, for a dromion solution, $p$ and $q$ must have opposite signs.}

For $q=0$ in (\ref{exc1_u,v,s}), setting $t_n=0$ for $n>2$, we have
\begin{gather}
   u = c_2   p^2   \big( c_1   p + c_4   p   e^{-2 p t_1}
      + e^{-p t_1}   ( p^2 e^{-p(x+p t_2)} + (c_1 c_4 - c_2 c_3)   e^{p(x+pt_2)} \big)^{-1}   ,
             \nonumber \\
   v = c_3   p^2   \big( c_4   p + c_1   p   e^{2 p t_1}
      + e^{p t_1}   ( p^2 e^{-p(x+p t_2)} + (c_1 c_4 - c_2 c_3)   e^{p(x+p t_2)} \big)^{-1}   .
       \label{c1ex1_q=0}
\end{gather}
If $c_1 c_4 > c_2 c_3$ and $c_1   p >0$, $c_4   p>0$ (or $c_1 c_4 < c_2 c_3$ and $c_1   p <0$, $c_4   p<0$),
these functions obviously represent wedge-shaped kinks in the $xt_1$-plane, see also Fig.~\ref{fig:kink2dromion}.
Choosing $p>0$ and switching on a negative $q$, these wedges become localized and, around certain
negative values of~$q$, then take the dromion form.

\begin{figure}[t]
\centerline{\includegraphics[width=16cm]{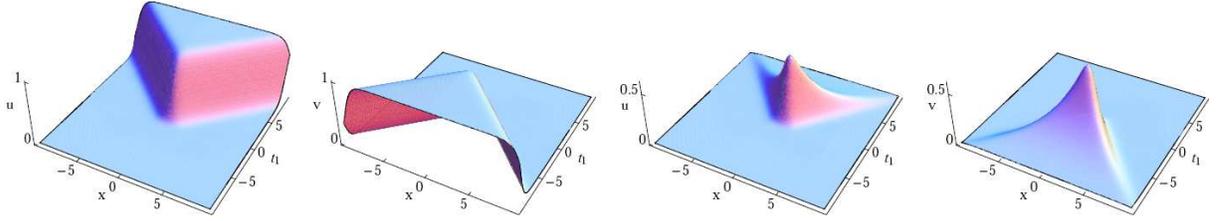}}

\caption{Solutions of (\ref{ex1_u,v_eqs}) and (\ref{ex1_s,r_eqs}) at $t_2=0$, given by
(\ref{c1ex1_phi}) with $p=2$, $c_1=c_4=1$, $c_2=c_3=1/2$.
The f\/irst two plots, where $q=0$, show kinks (see (\ref{c1ex1_q=0})). In the last two plots,
where $q=-1/5$, these become exponentially localized structures.
   \label{fig:kink2dromion} }
\end{figure}

Fig.~\ref{fig:2dromions} shows plots of a two-dromion solution determined by (\ref{case1_L,R}) with
$L=\mathrm{diag}(3   I_2, 2  I_2)$, $R=\mathrm{diag}(-2   I_2, -(3/2) I_2)$, and\footnote{If all
lower-diagonal entries of $C$ are zero, we obtain $v=0$ and thus a solution of the linear equations
(\ref{u,s_eqs_v=0}). The plots are surprisingly insensitive with respect to changes in this range
of parameters, as long as the of\/f-diagonal entries in a diagonal block are not all close to zero and
the determinant of the block is not close to zero. }
\begin{gather*}
   C = \left( \begin{array}{cccc} 1 & \frac{1}{2} & 0 & 0 \\
                                  0 & 1 & 0 & 0 \\
                                  0 & 0 & 2 & 1 \\
                                  0 & 0 & 5 & 3
                \end{array} \right)    .   
\end{gather*}
The diagonal $2 \times 2$ blocks of these matrices correspond to matrix data of single dromions.
Such a~superposition is obtained for any two solutions, provided that of\/f-diagonal blocks of
the matrix~$K$ exist such that (\ref{RK-KL=VU^T}) can be satisf\/ied. This is so in the restricted case
considered above (see~(\ref{case1_L,R}) and (\ref{case1_ex_K})), which in particular leads to
multi-dromion solutions.
Introducing non-zero constants in the of\/f-diagonal blocks of $C$, leads to solutions
with more complicated behaviour.

\begin{figure}[t]
\centerline{\includegraphics[width=13cm]{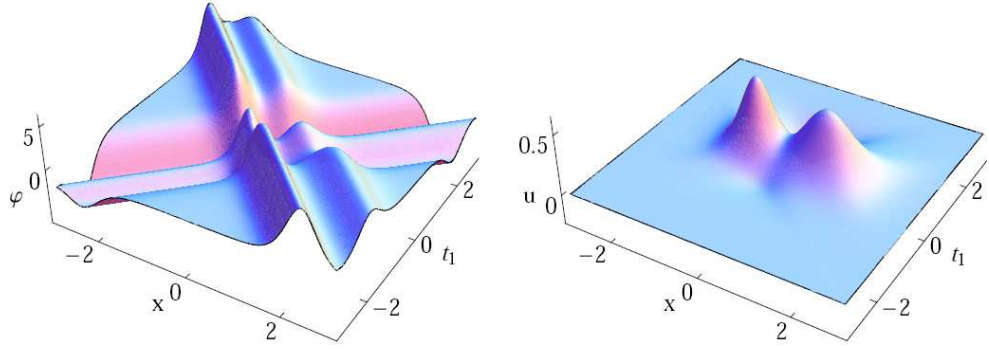}}

\caption{A two-dromion solution of (\ref{ex1_u,v_eqs}) and (\ref{ex1_s,r_eqs}) at $t_2=1$,
determined by the data specif\/ied in Example~\ref{ex:c1-ex1}.
The left plot shows $\varphi = s_{t_1} = \tr(\phi)_{t_1}$.
   \label{fig:2dromions} }
\end{figure}

Setting $t_n =0$, $n>2$, the transition to the Davey--Stewartson system (\ref{DS}) with
$\kappa = 1$, which is the DS-I case, is given by the transformation of independent variables
\[
    x = \frac{1+\imag}{2 \sqrt{2}} (z - \imag   y)   , \qquad
    t_1 = -\frac{1-\imag}{2 \sqrt{2}} (z + \imag   y)  , \qquad
    t_2 = - \imag   t   .
\]
The dependent variables are $u$ and $\rho$, the latter given by (\ref{DS_rho}).
We have to take the additional constraint $v = \imag   \epsilon   \bar{u}$
into account (see Example~\ref{ex:ds}). One recovers a DS-I dromion
within the class of solutions restricted by $q_1 = \bar{p}_1$, $q_2 = -\bar{p}_2$, $c_1$
imaginary and $c_4$ real, and $c_3 = \pm \imag \, \bar{c}_2$ with sign corresponding
to $\epsilon= \pm 1$. Fig.~\ref{fig:DSI_dromion} shows an example.

\begin{figure}[t]
\centerline{\includegraphics[width=13cm]{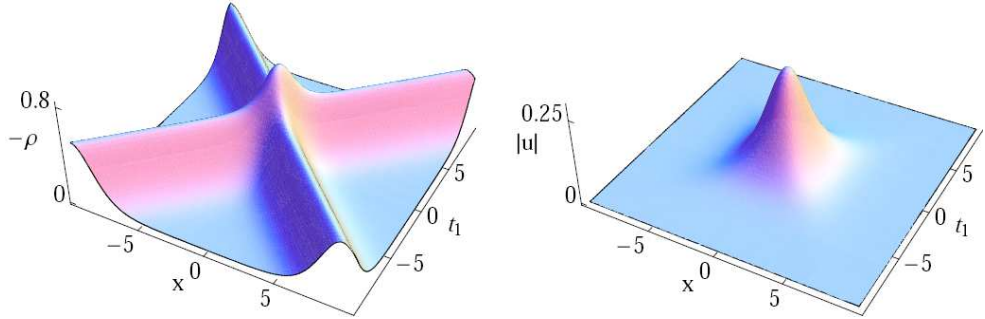}}

\caption{A dromion solution of the Davey--Stewartson-I equation with $\epsilon=-1$ at $t_2=0$, see
the end of Example~\ref{ex:c1-ex1}. Here we chose
$p_1 = p_2 = (1+\imag)/\sqrt{2}$, $q_1 = -q_2 = (1-\imag)/\sqrt{2}$, $c_1 = -c_2 = \imag   \sqrt{2}$,
$c_3 = c_4 = \sqrt{2}$. Here $\rho$ is given by (\ref{DS_rho}).
   \label{fig:DSI_dromion} }
\end{figure}

\end{example}

\noindent
\textbf{Case 2.} Let $M=N$, $R=L$, $S=0$, and
\begin{gather}
     Q = J + [L , K ]     ,   \label{Q=J+[L,K]}
\end{gather}
with constant $N \times N$ matrices $K$ and $J$ (over $\cA$) that commute with all $B \in \cB$.
Furthermore, $J$ has to commute with $L$, i.e.\ $[J,L]=0$. Then
\begin{gather}
   \Phi = e^{\xi(x,\mathbf{t};L,B)}   C   e^{-\xi(x,\mathbf{t};L,B)}
        \big(I_N + (\xi'(x,\mathbf{t};L,B)   J - K)   e^{\xi(x,\mathbf{t};L,B)}   C
         e^{-\xi(x,\mathbf{t};L,B)} \big)^{-1}    ,   \label{Phi-type2a}
\end{gather}
where $C$ is an arbitrary constant $N \times N$ matrix (with entries in $\cA$) and
\begin{gather*}
     \xi'(x,\mathbf{t};L,B) := x + \sum_{B \in \cB} \sum_{n=1}^\infty n   t_{B,n}   L^{n-1}   B^n    .
\end{gather*}
If also $Q=V U^\intercal$ with vectors $U$, $V$ that commute with all $B \in \cB$, then $\phi = U^\intercal \Phi V$
solves the mcKP hierarchy in $\cA$. It remains to solve
\begin{gather}
     J + [L , K ] = V U^\intercal   .   \label{J+[L,K]=VU^T}
\end{gather}
A natural choice for $J$ is the unit matrix $I_N$, but there are others. (\ref{Phi-type2a}) can also be written as
\begin{gather}
   \Phi = \big( e^{\xi(x,\mathbf{t};L,B)}   \tilde{C}   e^{-\xi(x,\mathbf{t};L,B)} + \xi'(x,\mathbf{t};L,B)   J - K
          \big)^{-1}     \label{Phi-type2b}
\end{gather}
with an arbitrary constant $N \times N$ matrix $\tilde{C}$.\footnote{In the transition from (\ref{Phi-type2a})
to (\ref{Phi-type2b}), one assumes that $C$ is invertible with inverse $\tilde{C}$. But $\tilde{C}$ need
not be invertible in order that (\ref{Phi-type2b}) determines a solution of the mcKP$_Q$ hierarchy.}
If $\tilde{C}$ is chosen such that it commutes with~$L$ and~$B$, then~$\Phi$ and the corresponding solution
$\phi$ of the mcKP hierarchy are purely \emph{rational} functions of the independent variables.
A localized solution of this kind, hence with rational decay, is usually called a ``lump''.
The following example in particular demonstrates that there can be weaker conditions that
lead to solutions with rational decay.

\begin{example}
Choosing $B = \mathrm{diag}(1,-1)$, $L = \mathrm{diag}(p_1,p_2)$ and $Q = I_2$, (\ref{Q=J+[L,K]}) is solved
by $K = \mathrm{diag}(k_1,k_2)$. Expressing $C$ again as in (\ref{Phi0}), we f\/ind
\[
    e^{\xi(x,\mathbf{t};L,B)}   C   e^{-\xi(x,\mathbf{t};L,B)}
  = \left( \begin{array}{cc} c_1 & c_2   e^{\xi_+(x,\mathbf{t};p_1)-\xi_-(x,\mathbf{t};p_2)} \\
          c_3   e^{-\xi_+(x,\mathbf{t};p_1)+\xi_-(x,\mathbf{t};p_2)} &
             c_4  \end{array} \right)   ,
\]
with $\xi_{\pm}(x,\mathbf{t};p)$ def\/ined in (\ref{xi_pm}),
and $\xi'(x,\mathbf{t};L,B) = \mathrm{diag}( \xi'_+(x,\mathbf{t};p_1) , \xi'_-(x,\mathbf{t};p_2) )$, where
\[
    \xi'_{\pm}(x,\mathbf{t};p) := x + \sum_{n=1}^\infty 2n   t_{2n} p^{2n-1}
                     \pm \sum_{n=0}^\infty (2n+1)   t_{2n+1} p^{2n}   .
\]
Then (\ref{Phi-type2a}) leads to the following solution of (\ref{Beq-la0mu1}), with $B = \mathrm{diag}(1,-1)$,
and its hierarchy,
\[
   \phi(x,\mathbf{t}) = \frac{1}{\mathcal{D}(x,\mathbf{t})} \left( \begin{array}{cc}
     c_1 + d   ( \xi'_-(x,\mathbf{t};p_2) - k_2 ) & c_2   e^{\xi_+(x,\mathbf{t};p_1)-\xi_-(x,\mathbf{t};p_2)}  \\
     c_3   e^{-\xi_+(x,\mathbf{t};p_1)+\xi_-(x,\mathbf{t};p_2)} & c_4 + d   ( \xi'_+(x,\mathbf{t};p_1) - k_1 )
        \end{array} \right)   ,
\]
where $d := c_1 c_4 - c_2 c_3$ and
\begin{gather*}
    \mathcal{D}(x,\mathbf{t})  :=  1 + c_1   (\xi'_+(x,\mathbf{t};p_1) - k_1)
                    + c_4   (\xi'_-(x,\mathbf{t};p_2) - k_2) \nonumber \\
             \phantom{\mathcal{D}(x,\mathbf{t})  :=}{}   + d   (\xi'_+(x,\mathbf{t};p_1)-k_1)(\xi'_-(x,\mathbf{t};p_2) - k_2)  .
\end{gather*}
The components $u = \phi_{1,2}$ and $v = \phi_{2,1}$, together with
\[
    s = \tr(\phi)
      = \frac{d}{\mathcal{D}} \big( \xi'_+(x,\mathbf{t};p_1) + \xi'_-(x,\mathbf{t};p_2) + c \big)  ,
\]
with a constant $c$, thus solve the system (\ref{ex1_u,v_eqs}), (\ref{ex1_s,r_eqs}). For $c_3=0$,
this determines a solution of the linear equations (\ref{u,s_eqs_v=0}).

The transition to the Davey--Stewartson system (\ref{DS}) with $\kappa = \imag$, which is the
DS-II case, involves the transformation of independent variables
\[
    x = \frac{1+\imag}{2 \sqrt{2}} (y + z)   , \qquad
    t_1 = \frac{1-\imag}{2 \sqrt{2}} (y - z)   , \qquad
    t_2 = - \imag   t   .
\]
We set $t_n =0$, $n>2$, in the following. The lump solution of the DS-II equation \cite{Sats+Ablo79,NakaA82PLA,NakaA82JMP}
(see also \cite{Foka+Ablo83,APP89,Peli95DS,Mana+Sant97}) is obtained as follows from
the above formula. Besides taking account of the constraint $v = \imag   \epsilon   \bar{u}$,
we have to arrange in particular that the exponential in $u$ becomes a phase factor
(up to some constant factor), i.e.\ the real part of its exponent has to be constant. This requires setting
$p_2 = -\imag   \bar{p}_1$.
It is more tricky to f\/ind conditions on the remaining parameters such that $\mathcal{D} \neq 0$ for all $y$, $z$, $t$,
so that the solution is regular. Choosing
\[
    c_4 = - \imag   \bar{c}_1   , \qquad
    c_3 = \imag \, \bar{c}_2   , \qquad
    k_2 = - \imag   \bar{k}_1   ,
\]
and renaming $k_1$, $p_1$ to $k$, $p$, we f\/ind that
\[
  \mathcal{D} = \frac{1}{4} ( |c_1|^2 + 2 |c_2|^2 ) \big( (y-y_0 - v_y   t)^2 + (z-z_0 - v_z   t)^2 \big)
                 + \frac{2 |c_2|^2}{|c_1|^2 + 2 |c_2|^2}   ,
\]
where $v_y = -2 \, \mathrm{Im}(p)$, $v_z = 2 \, \mathrm{Re}(p)$, and
\[
   y_0 = \mathrm{Re}(k) - \frac{ 2 \, \mathrm{Re}(c_1) }{|c_1|^2 + 2 |c_2|^2}
           , \qquad
   z_0 = \mathrm{Im}(k) + \frac{ 2 \, \mathrm{Im}(c_1) }{|c_1|^2 + 2 |c_2|^2}
           .
\]
The resulting DS-II solution
\begin{gather*}
    u = \frac{c_2}{\mathcal{D}} \, e^{\imag ( \mathrm{Im}(p)   y + \mathrm{Re}(p)   z
          - \mathrm{Re}(p^2)   t ) }   , \\
   \rho = \frac{1}{4 \mathcal{D}^2} \, \big( (|c_1|^2 + 2 |c_2|^2 )^2
      \left( (z-z_0 - v_z   t)^2 - (y-y_0 - v_y   t)^2  \right) - 8   |c_2|^2 \big)   ,
\end{gather*}
with $\rho$ def\/ined in (\ref{DS_rho}) and $\epsilon=1$, is regular whenever
$c_2 \neq 0$ and reproduces a well-known lump solution.

Again, (lump) solutions can be superposed by taking matrix data of (lump) solutions as diagonal
blocks of larger matrices $L$ and $C$. It then essentially remains to determine the of\/f-diagonal
blocks of the new matrix $K$ so that (\ref{J+[L,K]=VU^T}) holds.
\end{example}

\section{The matrix Riccati system associated\\ with the multicomponent KP hierarchy}
\label{section:mcRiccati}

Writing
\[
  H^n =: \left(\begin{array}{cc} R_n & Q_n \\ S_n & L_n \end{array}\right)
          ,
\]
the matrix linear system (\ref{Z_x=HZ}), (\ref{Z_tBn=(BH)^nZ}) implies the
\emph{matrix Riccati system}
\begin{gather}
    \Phi_x = S + L   \Phi - \Phi   R - \Phi   Q   \Phi   , \label{Riccati_x} \\
    \Phi_{t_{B,n}} =  B^n S_n +  B^n L_n \Phi - \Phi B^n R_n
                       - \Phi B^n Q_n \Phi   .  \label{Riccati_t_Bn}
\end{gather}
The two solution families presented in Section~\ref{section:mcKPsol} solve this matrix Riccati system,
with the respective conditions imposed on the matrices $L$, $Q$, $R$, $S$.

Abstracting from matrices and thinking of $L$, $R$, $S$ as (noncommutative) algebraic objects, their elimination
from the above system leads to the mcKP hierarchy with product modif\/ied by~$Q$ (cf.~\cite{DMH07def}).
To some extent the above Riccati system thus expresses the mcKP hierarchy as a hierarchy of
ordinary dif\/ferential equations.

Finite-size matrix Riccati equations, in particular with constant
coef\/f\/icient matrices as above, were discussed in a context related to integrable systems
already long ago (see e.g.~\cite{Wint83,dORW87}), but apparently not in the context of the
KP hierarchy. A special inf\/inite-size matrix Riccati system involving the shift operator in inf\/inite
dimensions appeared, however, in the framework of the Sato theory (see e.g.~\cite{DNS89,Taka89}).
In the one-component case with $B=I$, the above Riccati system, with suitable conditions imposed on
the coef\/f\/icient matrices, also generates solutions of the BKP and the CKP hierarchy \cite{DMH08oddKP}.
The Riccati system indeed generates solutions of various integrable systems and therefore
deserves to be studied in its own right.

\begin{remark}
\label{rem:Ricc_red}
For f\/ixed $r \in \mathbb{N}$, $r>1$, and for some f\/ixed $B \in \cB$, let us consider the condition
\begin{gather*}
    (H B)^r   Z_0 = Z_0   P   ,  
\end{gather*}
with an $N \times N$ matrix $P$ (over $\cA$). For the solution (\ref{Zsol}) of the linear matrix system
(\ref{Z_tBn=(BH)^nZ}), this implies $(H B)^{n r}   Z = Z  P^n$ for $n \in \mathbb{N}$,
hence $B^{n r} (R_{nr} X + Q_{nr} Y) = X P^n$ and $B^{n r} (S_{nr} X + L_{nr} Y) = Y P^n$,
and thus the algebraic Riccati equations
$B^{n r} (S_{nr} + L_{nr} \Phi) = Y P^n X^{-1}
  = \Phi   X P^n   X^{-1}
  = \Phi   B^{n r} ( R_{nr} + Q_{nr} \Phi )$.
The corresponding equations (\ref{Riccati_t_Bn}) of the Riccati system then read
\[
    \Phi_{t_{B,nr}} = B^{nr} ( S_{nr} +   L_{nr} \Phi ) - \Phi B^{nr} ( R_{nr} - Q_{nr} \Phi )
                    = 0,  \qquad    n=1,2,\ldots   .
\]
Hence $\phi$ solves the $(r,B)$-\emph{reduction}, i.e.\ the $r$-\emph{reduction} (multicomponent
version of $r$th Gelfand--Dickey hierarchy) with respect to $B$.
\end{remark}

\section{Conclusions}
\label{section:concl}

Any solution of a multicomponent Burgers (mcBurgers) hierarchy
is a solution of the corresponding multicomponent KP (mcKP) hierarchy.
Furthermore, there is a functional equation that determines the
mcKP hierarchy and has the form of an \emph{inhomogeneous}
mcBurgers hierar\-chy functional equation. We have also shown that the
mcKP linear system is equivalent to a~mcBurgers hierar\-chy, where the dependent
variable has the structure of an inf\/inite Frobenius companion matrix (which in
particular makes contact with~\cite{Zenc+Sant08}).

Moreover, we have shown how solutions of a mcKP hierarchy are obtained from solutions
of a~multicomponent linear heat hierarchy via a generalized Cole--Hopf
transformation. An important subcase generates solutions of a mcKP hierarchy
from solutions of a matrix linear system and we presented some explicit solution
formulae. They comprise in particular Davey--Stewartson dromions and lump solutions.

There is certainly a lot more to be (re)discovered using the rather simple but quite general
method in Section~\ref{section:mcKPsol} to construct exact solutions, but we are far from a
systematic way to explore the properties of solutions obtained in this way.
Furthermore, we have stressed the role of a~matrix Riccati hierarchy in this context.

\pdfbookmark[1]{References}{ref}

\LastPageEnding

\end{document}